\documentclass[preprint]{cimento} %for arXiv

\pdfoutput=1
\usepackage{epsfig,sint,cite,amsfonts,amssymb,amsmath,bbold}
\usepackage[font=small,labelfont={sf,bf}]{caption}
\usepackage{subcaption}
\usepackage[T1]{fontenc}
\usepackage{tikz}
\usetikzlibrary{decorations.pathreplacing}
\usetikzlibrary{patterns} %per usare i retini
\usepackage[english]{babel}
\usepackage{simplewick}
\usepackage{dsfont}
\usepackage[utf8]{inputenc}
\usepackage{siunitx}
\usepackage{physics}
\AtBeginDocument{\RenewCommandCopy\qty\SI}
\usepackage{wrapfig}
\usepackage{slashed}

\newcommand{\be}{\begin{equation}}
\newcommand{\ee}{\end{equation}}
\newcommand{\bea}{\begin{eqnarray}}
\newcommand{\eea}{\end{eqnarray}}
\newcommand{\bes}{\begin{eqnarray}}
\newcommand{\ees}{\end{eqnarray}}
\newcommand{\ba}{\begin{array}}
\newcommand{\ea}{\end{array}}

\def\diracstar#1#2{
    \setbox0=\hbox{$\gamma$}\setbox1=\hbox{$\gamma_{#1}$}
    \gamma_{#1}\kern-\wd1\kern\wd0
    \smash{\raise4.5pt\hbox{$\scriptstyle#2$}}}

\usetikzlibrary{decorations.markings}
\usetikzlibrary{svg.path}
\usetikzlibrary{shapes,fit}

\renewcommand{\i}{\mathrm{i}}
\def\tmin{t_\mathrm{min}}
\def\tmax{t_\mathrm{max}}
\def\syst{\sigma_\mathrm{syst}}
\def\stat{\sigma_\mathrm{stat}}

\title{On the prediction of spectral densities from Lattice QCD}
\author{M.~Bruno\from{UNIMIB-INFN}, L.~Giusti\from{UNIMIB-INFN} \atque M.~Saccardi\from{UNIMIB-INFN}\thanks{Speaker}}
\instlist{\inst{UNIMIB-INFN} Dipartimento di Fisica, Universit\`a di Milano--Bicocca, and INFN, sezione di Milano--Bicocca \\ Piazza della Scienza 3, I-20126 Milano, Italy}

\begin{document}

\maketitle

\begin{abstract}
Hadronic spectral densities play a pivotal role in particle physics, a prime example being the R-ratio defined from electron-positron scattering into hadrons. To predict them from first principles using Lattice QCD, we face a numerically ill-posed inverse problem, due to the Euclidean signature adopted in practical simulations. Here we present a recent numerical analysis of the vector isovector spectral density extracted using the multi-level algorithm (recently extended also to the case of dynamical fermions) and discuss its implications.
\end{abstract}

\section{Introduction and motivations\label{sec:intro}}
Hadronic spectral densities are directly involved in the computation of many physically relevant quantities, \textit{e.g.} the anomalous magnetic moment of the muon \cite{Aoyama_2020} or the $V_{cb}$ element of the CKM matrix \cite{Amhis_2021,Gambino_2019}. Lattice QCD provides a framework for their non-perturbative prediction from first principles, with simulations on a discretized, finite lattice with Euclidean metric. Spectral densities $\rho(E)$ are related to correlation functions $C(t)$ along the Euclidean time axis $t$ as
\be \label{eq:rho_def}
    C(t) = \int_{E_0}^\infty dE \, \rho(E) b_t(E), \quad b_t(E) = e^{-tE}, \quad 0 \le E_0 < E_{\text{thr}} \,;
\ee
for theories with a mass gap $E_{\text{thr}}>0$. To reconstruct $\rho$ from measurements of $C$, the textbook Bromwich integral in the complex plane to solve the inverse Laplace transform is impractical due to the (discrete) Euclidean times at which $C$ is known in lattice calculations. In this work, we modify the Backus-Gilbert original method~\cite{Backus_Gilbert,Hansen_2017,Hansen_2019}, focusing on identifying the ideal balance between statistical and systematic errors in the reconstruction. Additionally, we present preliminary results for the isovector vector spectral density smeared with a Gaussian kernel with width $\sigma=M_\pi$, obtained from the corresponding Euclidean correlators estimated with high accuracy in Ref.~\cite{DallaBrida:2020cik} using the multi-level algorithm. The calculation is performed at a single lattice spacing with two degenerate dynamical fermions, tuned such that $M_\pi \simeq 270 ~\mathrm{MeV}$.

% analyze data generated using the multi-level algorithm on a gauge ensemble at the unphysical value of $M_\pi=270$ MeV~\cite{DallaBrida:2020cik}, presenting a first, preliminary result for the isovector vector spectral density smeared with a Gaussian kernel of width $\sigma=M_\pi=270$ MeV. 
% This shows the advantages of such algorithms in spectral reconstructions across a broad energy range.

% Maybe, include also BNL_2006,FNAL_2021 for g-2 reference, and Gambino_2020,Gambino_2022 for V_{cb}

\section{The inverse problem\label{sec:1}}
\begin{figure}[t!]
\centering
\begin{minipage}[t]{.49\textwidth}
    \includegraphics[width=.95\linewidth]{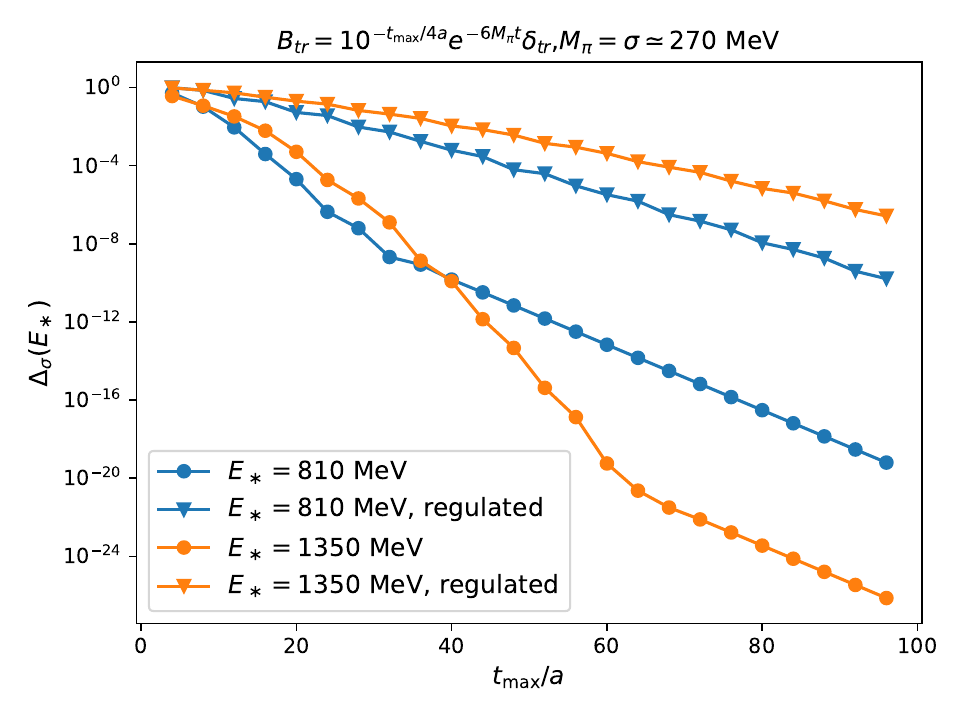}  
    \caption{Relative systematic error on the kernel reconstruction $\Delta_\sigma(E_\ast)$, see Eq.~(\ref{eq:Delta_sigma})}
    %\caption{$\left( \int_{E_0}^\infty dE \, \bar\delta_\sigma(E,E_\ast)^2 - C \right)/C, C = \int_{E_0}^\infty dE \, \delta_\sigma(E,E_\ast)^2 $}
    \label{Fig:tmax_error}
\end{minipage}
\hfill
\begin{minipage}[t]{.49\textwidth}
\centering
    \includegraphics[width=.95\linewidth]{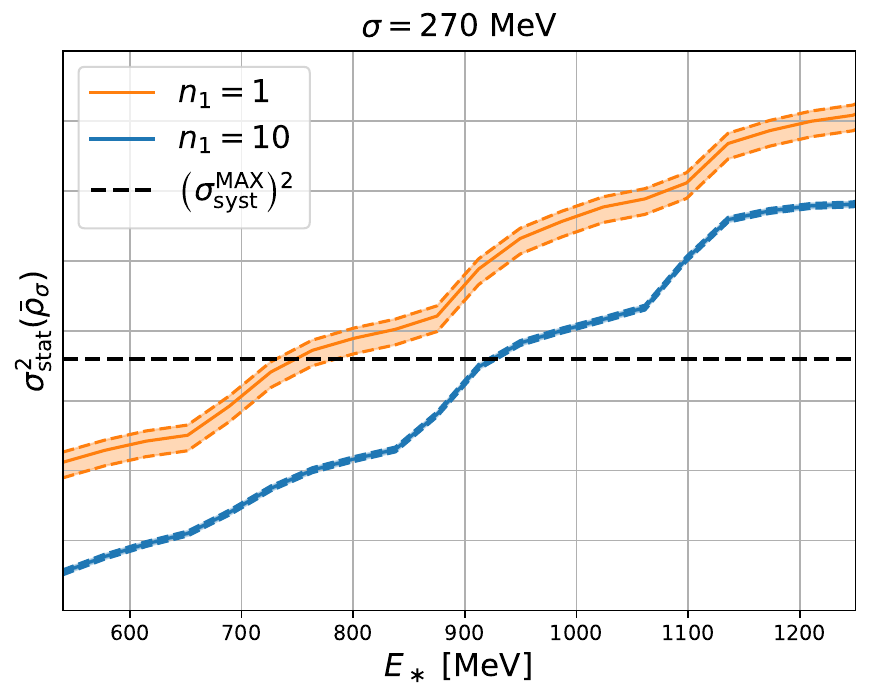}  
    \caption{Statistical errors on $\bar\rho_\sigma(E_\ast)$ at $\lambda_\ast: \syst^{\text{MAX}}(\lambda_\ast,E_\ast)=0.02$ $\forall \, E_\ast$ for $n_1=1,10$}
    \label{Fig:stat_errs_n1s}
\end{minipage}
\end{figure}

The extraction of smeared spectral densities from lattice correlation functions in Eq.~(\ref{eq:rho_def}) can involve different correlator-dependent choices of basis functions $b_t(E)$. In our case, $C(t)$ is the zero-momentum two-point function of two isovector vector currents and $b_t(E) = E^2 \left( e^{-tE} + e^{-(T-t)E} \right)$, with $T$ the extent of the time direction along which we impose (anti-)periodic boundary conditions for bosons (fermions). We are interested in extracting smeared spectral densities\footnote{Lattice calculations are performed in a finite box and we expect finite-volume effects to be of $O(e^{-\sigma L})$ for $\sigma \lesssim M_\pi$ \cite{Bruno_Hansen_2023}. Therefore, a non-zero smearing width is expected to be beneficial.}
\be
    \rho_\sigma(E_\ast) = \int_{E_0}^\infty dE \, \rho(E) \delta_\sigma(E,E_\ast), \quad \text{with } \delta_\sigma(E,E_\ast) = \frac{1}{\sqrt{2\pi\sigma^2}} e^{-(E-E_\ast)^2/2\sigma^2}
\ee
such that, as $\sigma\to0$, $\delta_\sigma(E,E_\ast)\to\delta(E-E_\ast)$ and we recover $\rho_\sigma(E_\ast)\to\rho(E_\ast)$. The kernel $\delta_\sigma$ is approximated with the basis functions as
\be
    \sum_{t=\tmin}^{\tmax} b_t(E) g_t(E_\ast) = \bar\delta_\sigma(E,E_\ast) \approx \delta_\sigma(E,E_\ast),
\ee
where we usually consider $\tmin=a$ and require $\tmax<T/2$, so that
\be \label{eq:bar_rho_sigma}
    \sum_{t=\tmin}^{\tmax} C(t) g_t(E_\ast) = \int_{E_0}^\infty dE \, \rho(E) \bar\delta_\sigma(E,E_\ast) \equiv \bar\rho_\sigma(E_\ast) \approx \rho_\sigma(E_\ast) \, .
\ee
The computation of the coefficients $\{g_t(E_\ast)\}_{t=\tmin}^{\tmax}$ is done by minimizing the error functional
\be
    \mathcal{W}[g] = \int_{E_0}^\infty dE \, \left[ \delta_\sigma(E,E_\ast) - \bar\delta_\sigma(E,E_\ast) \right]^2 + \sum_{t,r=\tmin}^{\tmax} g_t(E_\ast) B_{tr} g_r(E_\ast)
\ee
where the first term measures the systematic error on the kernel reconstruction and the second one acts as a regulator in the sense explained below. We obtain $g_t(E_\ast) = \left[ W^{-1} f(E_\ast) \right]_t$, where
\be
    W=A+B, \quad A_{tr} = \int_{E_0}^\infty dE \, b_t(E) b_r(E), \quad f_t(E_\ast) = \int_{E_0}^\infty dE \, b_t(E) \delta_\sigma(E,E_\ast)\, .
\ee
By setting $B=0$, the coefficients are expressed in terms of $A^{-1}$, where $A$, for the choice $b_t(E) = e^{-tE}$, is a Cauchy matrix directly related to the Hilbert matrix%\cite{Hilbert_1900}
, a notably ill-conditioned matrix%\cite{Schechter_1959}
. It is well-known, see \textit{e.g.} Ref.~\cite{Hansen_2019}, that this results in large oscillations in $g_t(E_\ast)$, and statistical fluctuations on $C(t)$ then induce large statistical errors on $\bar\rho_\sigma$. This is a manifestation of the ill-posed nature of the inverse Laplace transform applied to a finite set of discrete data with statistical error, the case of Lattice QCD. In order to mitigate this problem, we may regularize the near-zero eigenvalues of $A$ by setting $B\neq0$, adding yet another source of systematic error, besides the one due to $\tmax < \infty$. In \cite{Hansen_2017,Hansen_2019} it was proposed to consider $B$ to be proportional to the covariance matrix of data $\text{Cov}$. In principle, other types of regularizations that do not depend on data can be employed, \textit{e.g.} we can consider $B=\lambda \mathbf{1}$ (Tikhonov regularization) or $B_{tr} = \lambda e^{-Mt} \delta_{tr}$, where $\lambda, aM \in [0,\infty)$, so that $B \to \lambda \mathbf{1}$ for $aM\to 0$, and $B \to 0$ for $aM \to \infty$. In Fig.~\ref{Fig:tmax_error} we show how the (relative) systematic error on the kernel reconstruction
\be \label{eq:Delta_sigma}
    \Delta_\sigma(E_\ast) = \frac{\int_{E_0}^\infty dE \, \left[ \delta_\sigma(E,E_\ast) - \bar\delta_\sigma(E,E_\ast) \right]^2}{\int_{E_0}^\infty dE \, \delta_\sigma(E,E_\ast)^2} = 1 - \frac{\sum_{t,r=a}^{\tmax} f_t(E_\ast) [A^{-1}]_{tr} f_r(E_\ast)}{\int_{E_0}^\infty dE \, \delta_\sigma(E,E_\ast)^2}
\ee
decreases as we increase $\tmax$ or remove the regulator $B$. On the other hand, $B$ is needed to regularize $A$, taming the statistical fluctuations on $\bar\rho_\sigma$, an effect that is evident in Fig.~\ref{Fig:reconstructiona}. The regulator parameter $\lambda$ should then be tuned in order to minimize the statistical fluctuations $\stat$ while controlling the systematic errors $\syst$ on $\bar\rho_\sigma$, where we define
\be \label{eq:stat_and_syst}
    \stat^2(\lambda,E_\ast) = \sum_{t,r=\tmin}^{\tmax} g_t(E_\ast) \text{Cov}_{tr} g_r(E_\ast), \quad \syst(\lambda,E_\ast) = \vert \rho_\sigma(E_\ast) - \bar\rho_\sigma(E_\ast) \vert \, .
\ee
Since we do not have access to the true value $\rho_\sigma$, different estimates of $\syst$ can be given, see \textit{e.g.} in Refs.~\cite{Hansen_2019,Bulava_2021}. Here we use the following definition
\be \label{eq:syst_max}
    \syst(\lambda,E_\ast) \le \rho_{max} \int_{2 M_\pi}^{6 M_\pi} dE \, \left| \delta_\sigma(E,E_\ast) - \bar\delta_\sigma(E,E_\ast) \right| \equiv \syst^{\text{MAX}}(\lambda,E_\ast)
    % \\ \rho_{max} = \frac{R(s)_{max}}{12\pi^2 Z_V^2}, \quad R(s)_{max} \simeq R(s=M_\rho^2) \simeq 20, \, Z_V \simeq 0.75
\ee
which is an upper bound on the systematic error for $\rho_\sigma$ at small energies, if $\rho(E) \ge 0$ $\forall \, E$ and $\rho(E) \le \rho_{max} \equiv \max_E \rho(E)$ $\forall E \in [2 M_\pi, 6 M_\pi]$; preliminary results with $\lambda$ tuned so that $\stat(\lambda_\ast,E_\ast) = \syst^{\text{MAX}}(\lambda_\ast,E_\ast)$ are presented in Sect.~\ref{sec:2}. A proper analysis without the restriction of the integral in Eq.~(\ref{eq:syst_max}) to $6 M_\pi$ is ongoing. A different strategy that is currently being explored consists in estimating
\be \label{eq:double_recon}
    \rho_\sigma(E_\ast) - \bar\rho_\sigma(E_\ast) = \int_{E_0}^{\infty} dE \, \left[ \delta_\sigma(E,E_\ast) - \bar\delta_\sigma(E,E_\ast) \right] \rho(E) \equiv \int_{E_0}^{\infty} dE \, \delta_\sigma^{(1)}(E,E_\ast) \rho(E)
\ee
by means of a second reconstruction with the smearing kernel $\delta_\sigma^{(1)}(E,E_\ast)$. %In principle, we could then define an improved estimator $\bar\rho^{(1)}_\sigma(E_\ast)$ for $\rho_\sigma$ by reabsorbing the estimate of the systematic error in the previous estimate of $\bar\rho_\sigma$. By iterating this procedure with a fixed regulator parameter for each reconstruction, we recover the non-regulated, numerically ill-defined result $\bar\rho^{(\infty)}_\sigma(E_\ast) = \sum_{t,r=\tmin}^{\tmax} C(t) [A^{-1}]_{tr} f_r(E_\ast)$.

\section{Multi-level algorithm and spectral densities\label{sec:2}}
\begin{figure}[t]
\centering
\begin{subfigure}[t]{.49\textwidth}
    \includegraphics[width=.95\linewidth]{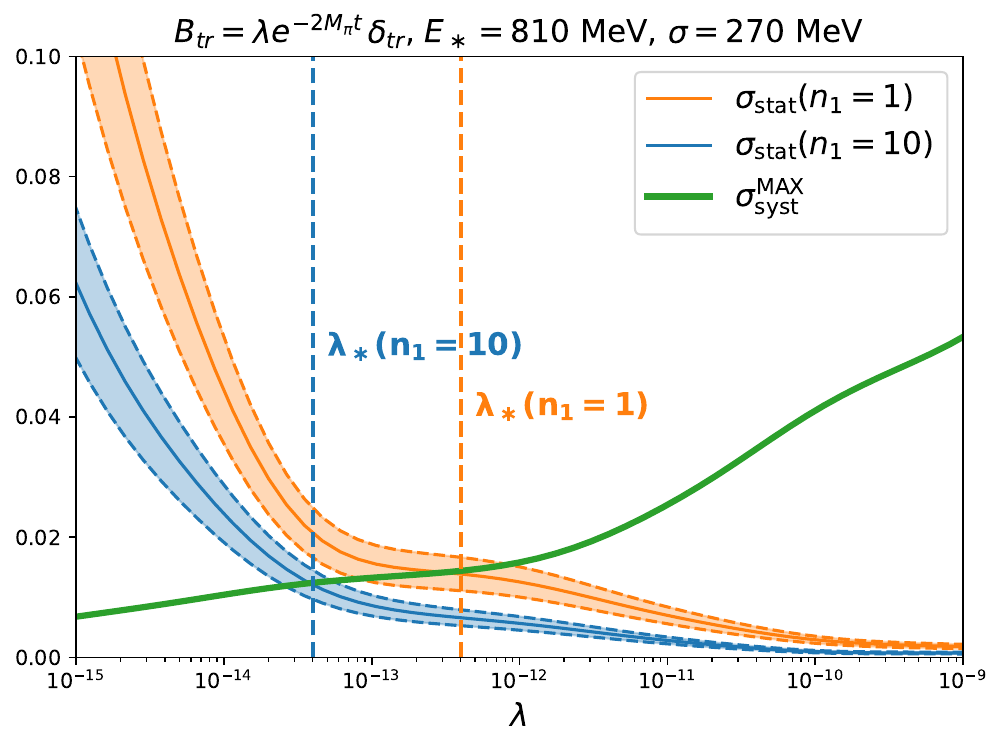}
    \caption{Tuning of $\lambda$}
    \label{Fig:reconstructiona}
\end{subfigure}
\hfill
\begin{subfigure}[t]{.49\textwidth}
    \includegraphics[width=.95\linewidth]{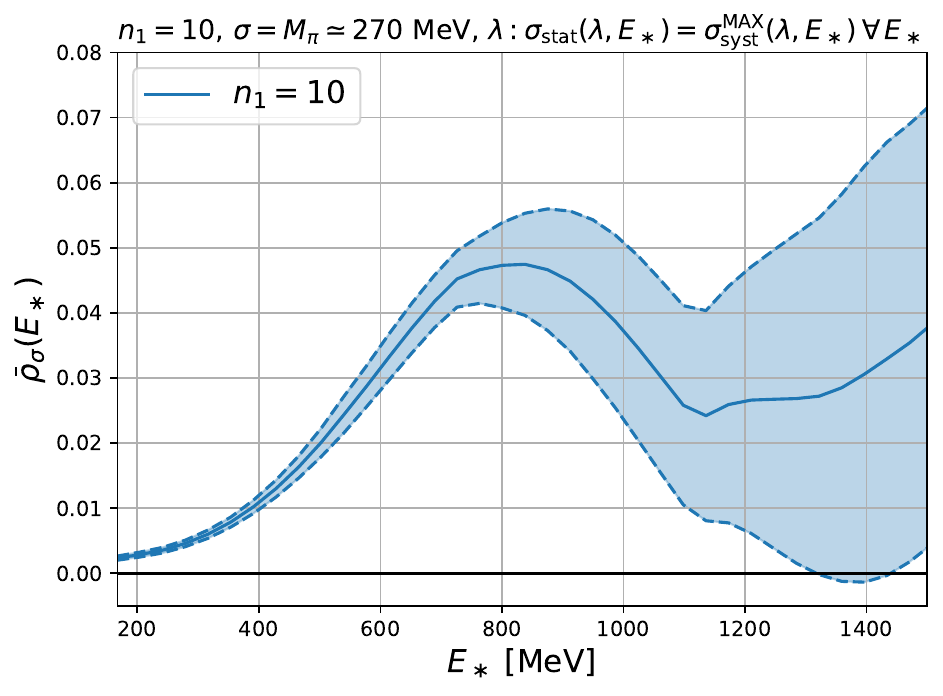}
    \caption{$\bar\rho_\sigma$}
    \label{Fig:reconstructionb}
\end{subfigure}
\caption{Reconstruction of $\bar\rho_\sigma(E_\ast)$ at $n_1=1,10$, tuning $\lambda: \stat(\lambda_\ast,E_\ast) = \syst^{\text{MAX}}(\lambda_\ast,E_\ast)$ $\forall \, E_\ast$}
\label{Fig:reconstruction}
\end{figure}

In this Section, we present preliminary numerical results for the extraction of smeared spectral densities from lattice correlation functions measured with the multi-level algorithm. In lattice calculations, many physically relevant observables suffer from an exponentially decreasing signal-to-noise ratio (StN) \cite{Parisi:1983ae%,Lepage:1989hd
}. If the action and the observable dependence on the field variables can be factorized \textit{e.g.} in two local domains as in \cite{Ce:2016idq,Ce:2016ajy,Ce:2017ndt,DallaBrida:2020cik}, we can alternate $n_0$ independent global (level-0) updates of the field configuration with $n_1$ (level-1) updates of the local regions only. %The combination of different configurations generated for each local domain results in an exponential gain in the StN:
Combining configurations from different local domains results in an exponential gain in the StN:
\be
    \frac{C(t)}{\stat(t)} \propto n_0 \cdot n_1^2 \, e^{-(M_\rho-2M_\pi)t} \, .
\ee
The factorization is straightforward for pure gauge theories \cite{Luscher:2001up}, and tremendous theoretical and numerical progress has been recently made for theories with dynamical fermions \cite{Luscher:2001up,DellaMorte:2007zz,DellaMorte:2008jd,Ce:2016idq,Ce:2016ajy,Ce:2017ndt,GIUSTI2022137103}.

The configurations were generated in Ref.~\cite{DallaBrida:2020cik} for $N_\mathrm{f}=2$ dynamical $O(a)$-improved Wilson fermions on a $96 \times 48^3$ lattice with spacing $a = 0.065$ fm and pion mass $M_\pi \simeq 270$ MeV, so that $M_\pi L \simeq 4.3$. The local, unimproved discretization of the vector current was used, with the value of the renormalization factor $Z_V$ at $\beta=5.3$ taken from Ref.~\cite{Dalla_Brida_2019}. \\
First, we study the extraction of $\bar\rho_\sigma(E_\ast)$ at a fixed value of $\syst^{\text{MAX}}=0.02$ $\forall \, E_\ast$ by tuning $\lambda$ accordingly. In Fig.~\ref{Fig:stat_errs_n1s} we show the statistical errors for $n_1=1$ and $n_1=10$ with a Gaussian kernel of width $\sigma=M_\pi\simeq270$ MeV and a regulator of the form $B_{tr} = \lambda e^{-2M_\pi t} \delta_{tr}$. At a number of configurations which is $n_1=10$ times larger, at all energies we achieve a gain in the statistical errors of $\bar\rho_\sigma(E_\ast)$ which ranges between $25$ and $50$. The difference from the ideal $n_1^2=100$ scaling was already observed in Ref.~\cite{DallaBrida:2020cik} for the variance of $C(t)$ and is compatible with the presence of a residual correlation among level-1 configurations. Thus, at fixed systematic error, the multi-level scaling of the statistical error proves to be advantageous when compared to classical, one-level ($n_1=1$) algorithms at the energies we probe. From the same figure we note that, in order to minimize the statistical fluctuations while keeping the systematic errors under control across all energy levels, we should increase or decrease $\lambda$ at lower or higher energies respectively. \\
Following this last observation, in Fig.~\ref{Fig:reconstructiona} we tune $\lambda$ so that $\stat(\lambda_\ast,E_\ast) = \syst^{\text{MAX}}(\lambda_\ast,E_\ast)$ for a chosen $E_\ast$. We note that $n_1=1$ data require a larger value of $\lambda$, thus increasing the overall error. By repeating this tuning for a wider energy range, from our best data-set at $n_1=10$ we obtain the preliminary results shown in Fig.~\ref{Fig:reconstructionb} for the noticeably small value of the kernel width $\sigma=M_\pi\simeq270$ MeV. Similarly to other studies \cite{Hansen_2019,Bulava_2021}, we observe an increase of the statistical error with the energy, but with our multi-level data we observe the presence of a peak, which we fit with a (smeared) Breit-Wigner ansatz; our preliminary results are compatible with the presence of a resonance but further refinements are needed to overcome the present limitations of Eq.~(\ref{eq:syst_max}) and to assess remaining systematic effects, including \textit{e.g.} the dependence on the unphysical pion mass. 

%Besides showing a rapid loss of signal, the reconstruction at $n_1=1$ is qualitatively different from that at $n_1=10$, which is the only one clearly revealing the underlying structure, \textit{e.g.} the presence of a resonance peak.

\section{Conclusions and outlook\label{sec:concl}}
The multi-level sampling strategy has proved to be effective for the reconstruction of smeared hadronic spectral densities at the energies that we probed. A detailed study of the scaling properties of the error of $\rho_\sigma$ and the study of other channels, such as the isovector axial and isoscalar vector channels, is deferred to future publications. Here we presented some preliminary results with a conservative estimate of the systematic error (especially at lower energies), showing promising results for $\rho_\sigma$ with a kernel width $\sigma=M_\pi$. A large effort is underway to improve the definition and understanding of the systematic uncertainties associated to the extraction of smeared spectral densities. Besides the statistical and systematic errors examined here, finite-volume \cite{Bruno_Hansen_2023} and discretization effects will have to be taken into account as well. Finally we remark, as described in Ref.~\cite{Alexandrou_2022}, that at physical $M_\pi$ one can directly compare Lattice QCD results for $\bar\rho_\sigma$ with experimental hadronic spectral densities smeared with the same kernel.

\section{Acknowledgements}
We thank M. Dalla Brida, T. Harris and M. Pepe for data availability. This work is (partially) supported by ICSC – Centro Nazionale di Ricerca in High Performance Computing, Big Data and Quantum Computing, funded by European Union – NextGenerationEU. The research of M.B. is funded through the MUR program for young researchers ``Rita Levi Montalcini''.

%\section{Acknowledgements}
%We thank L. Giusti for the many relevant discussions on the topic and M. Dalla Brida, L. Giusti, T. Harris and M. Pepe for data availability. The research of M.B. is funded through the MUR program for young researchers ``Rita Levi Montalcini''.

%\bibliographystyle{JHEP}
%\bibliography{mb.bib}

\end{document}